\documentclass[letterpaper,twocolumn,showpacs,
prl
,superscriptaddress%
]{revtex4}

\usepackage{graphicx}
\usepackage{amssymb}
\usepackage{amsmath,amsfonts,latexsym}
\usepackage{array,tabularx}
\usepackage{dcolumn}               

\DeclareMathOperator{\Ryx}{\mathcal{R}y_{x}}
\DeclareMathOperator{\ex}{x}

\newcommand{\vect}[1]{{\mathbf #1}}
\newcommand{\Frac}[2]{\displaystyle\frac{#1}{#2}}

\begin{document}

\title{Thermodynamics and Excitations of Condensed Polaritons in
  Disordered Microcavities}

\author{F.~M.~Marchetti}
\affiliation{Cavendish Laboratory, University of Cambridge,
             Madingley Road, Cambridge CB3 0HE, UK}
\author{J.~Keeling}
\affiliation{Cavendish Laboratory, University of Cambridge,
             Madingley Road, Cambridge CB3 0HE, UK}
\author{M.~H.~Szyma{\'n}ska}
\affiliation{Clarendon Laboratory, Department of Physics, University of Oxford,
             Parks Road, Oxford, OX1 3PU, UK}
\author{P.~B.~Littlewood}
\affiliation{Cavendish Laboratory, University of Cambridge,
             Madingley Road, Cambridge CB3 0HE, UK}

\date{January 11, 2005}       

\begin{abstract}
  We study the thermodynamic condensation of microcavity polaritons
  using a realistic model of disorder in semiconductor quantum
  wells. This approach correctly describes the polariton inhomogeneous
  broadening in the low density limit, and treats scattering by
  disorder to all orders in the condensed regime. While the weak
  disorder changes the thermodynamic properties of the transition
  little, the effects of disorder in the condensed state are prominent
  in the excitations and can be seen in resonant Rayleigh scattering.
\end{abstract}


\maketitle

Considerable effort has been recently devoted to the realisation of a
Bose-Einstein condensate of polaritons in III-V and II-VI
semiconductor microcavities~\cite{lesidang,deng,richard}. The very
light mass of these composite bosonic particles promises relatively
high transition temperatures, establishing these systems as ideal
candidates for observing condensation. A significant challenge to the
realisation of an equilibrium condensate might be represented by the
short polariton lifetime (caused by the finite quality of cavity
mirrors) and by the suppression of thermalization processes by
acoustic phonons at small momenta -- the `bottleneck effect'.
However, very recent developments have suggested that, by positively
detuning the cavity energy above the exciton energy, and by increasing
the non-resonant pump power to increase particle-particle scattering,
thermalization can be dramatically amplified~\cite{private}. While
unambiguous evidence for equilibrium condensation still remains
uncertain, much progress has been achieved in this direction. This
includes the observation of a non-linear threshold behaviour in the
emission intensity under non-resonant pump, the decrease, above
threshold, of the second order coherence function~\cite{deng} together
with a characteristic change in the momentum space distribution and,
recently, interference patterns in far-field emission have been
measured~\cite{richard}.

Theoretical effort has also been devoted to predicting properties and
possible signatures of polariton
condensation~\cite{jonathan,bosonic}. In this Letter we consider how
disorder, through the distribution of excitonic energies and
oscillator strengths, affects such signatures. Even with the
sophisticated growth technologies used in current structures, the
presence of interface and alloy disorder induces noticeable
effects. Here, we will show that the response under resonant Rayleigh
scattering (RRS), the coherent scattering by the disorder of an
injected photon into directions other than its original direction,
provides a unique probe of the condensed regime. Already in the low
density (linear) regime, disorder determines the RRS
response~\cite{freixanet,langbein} and the inhomogeneous broadening of
the polariton photoluminescence (PL). Using a quantitatively accurate
model for the exciton disorder~\cite{runge_review,zimmermann} that has
already been well-tested in the linear regime of excitation, we
investigate the effects of disorder on an equilibrium polariton
condensation. We find that the thermodynamic properties of the
transition are weakly affected by small disorder, as
expected. However, at densities above the threshold expected for
condensation, the normal modes supported by the cavity change from the
lower and upper polariton modes to new collective
excitations~\cite{jonathan}. Accordingly, the response of the
condensate to an additional RRS probe also changes. In noticeable
contrast with the non-condensed regime, the frequency-resolved RRS
emission from the linear Goldstone mode exists both above and below
the chemical potential. Moreover, the spectrum exhibits features
directly related to the disordered quasi-particle spectrum. Here, in
analogy with the BCS theory, the quasi-particles are given by
`particle-hole' excitations (i.e. bound excitons) coupled to the
coherent photon field via the random, disorder dependent, oscillator
strength.

The linear response of a resonantly pumped polaritonic system to an
external disorder potential has been recently studied
in~\cite{carusotto}. In that work, in contrast with the case analysed
here, the coherence of the system is driven by the pump, and moreover
the effect of disorder is included at a perturbative level.

The influence of quantum well disorder on excitonic energies and
oscillator strengths has been much studied in the last two decades
(for an exhaustive discussion see, e.g.,~\cite{runge_review} and
Refs. therein). Here, we assume the disorder to be correlated on a
length scale $\ell_c$ larger than the exciton Bohr radius $a_{\ex} =
\epsilon/e^2 \mu_r$, where $\mu_r$ is the reduced mass (henceforth we
will set $\hbar = 1$). Accordingly, we factorise the in-plane relative
and centre of mass coordinates, $\Psi_{\alpha} (\vect{r}_e,
\vect{r}_h) \simeq \varphi_{1s} (r) \Phi_{\alpha} (\vect{R})$, and
focus on the excitonic centre of mass motion,
\begin{equation}
  \left[-\Frac{\nabla_{\vect{R}}^2}{2m_{\ex}} + V(\vect{R})\right]
  \Phi_{\alpha} (\vect{R}) = \varepsilon_{\alpha} \Phi_{\alpha}
  (\vect{R})\; ,
\label{eq:centr}
\end{equation}
where the energies are measured w.r.t. the exciton band edge
$E_{\ex}$, i.e., the band gap minus the exciton Rydberg $\Ryx =
(2\mu_r a_{\ex}^2)^{-1}$. The effective disorder potential $V
(\vect{R})$ represents the microscopic structural disorder averaged
over the electron-hole motion~\cite{zimmermann} and can be
approximated, e.g., with a Gaussian noise, $\langle V(\vect{R})
V(\vect{R}')\rangle = (\sigma^2 \ell_c^2/L^2)
\sum_{\vect{q}}^{1/\ell_c} e^{i\vect{q} \cdot (\vect{R} -
\vect{R}')}$, where $L^2$ is the quantisation area.

\begin{figure}
\begin{center}
\includegraphics[width=1\linewidth,angle=0]{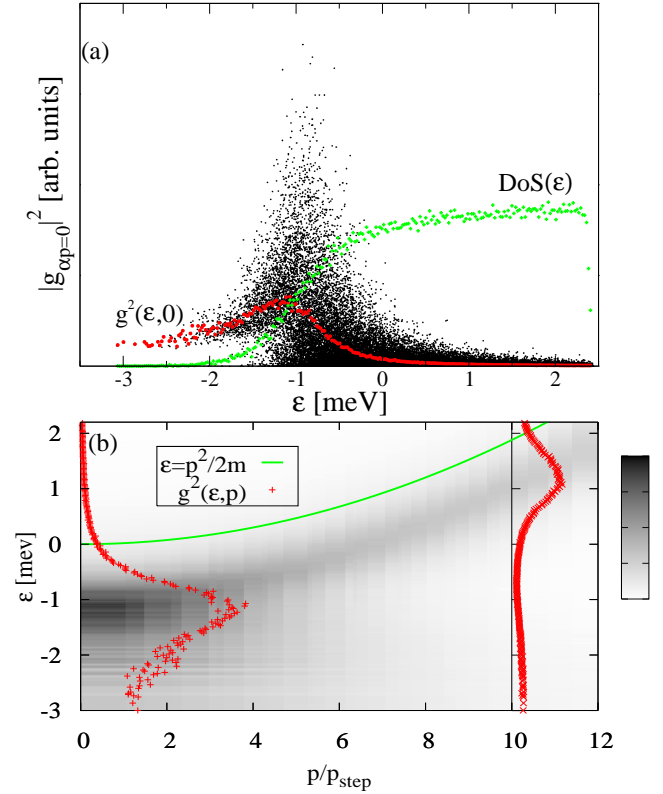}
\end{center}
\caption{\small (a) Plot of the squared coupling strength
         $|g_{\alpha,\vect{p}}|^2$ versus energy (160 realisations of
         disorder) for $\vect{p}=0$; (b) contourplot of the squared
         averaged oscillator strength $g^2 (\varepsilon,|\vect{p}|)$
         versus energy and momentum. The resolution in momentum,
         $p_{\text{step}} = 2\pi /L = 6.3 \times 10^{4}$cm$^{-1}$
         corresponds, for a cavity of $\omega_0=1.68$eV, to an angle
         of $\theta_{\text{step}} = \tan^{-1} (c
         p_{\text{step}}/\omega_0)= 36^{\circ}$. The free particle
         dispersion $|\vect{p}|^2/2m_{\ex}$ (solid line), and the
         squared averaged oscillator strength for two values of
         momenta, $|\vect{p}|=0$ and $|\vect{p}|=10 p_{\text{step}}$
         ($\theta = 82^{\circ}$) (plus symbols) are explicitly
         plotted.}
\label{fig:gepsi}
\end{figure}

Even though, in two dimensional non-interacting systems, all states
are localised by the disorder potential, the character of the
excitonic wavefunction changes significantly from below to above the
band edge $E_{\ex}$. The coupling strength of an exciton to light
changes accordingly. For dipole-allowed transitions, the exciton
oscillator strength $g_{\alpha,\vect{p}}$ is proportional to the
probability amplitude $\varphi_{1s} (0) \Phi_{\alpha,\vect{p}}$ of
finding an electron and a hole at the same position and with centre of
mass momentum equal to the photon momentum
$\vect{p}$~\cite{runge_review,zimmermann},
\begin{equation}
  g_{\alpha ,\vect{p}} = e d_{ab} \sqrt{\Frac{2\pi
  \omega_{\vect{p}}}{\epsilon L_w}} \varphi_{1s} (0)
  \Phi_{\alpha,\vect{p}}\; ,
\label{eq:oscil}
\end{equation}
where $d_{ab}$ is the dipole matrix element. 
Here, we solve Eq.~\eqref{eq:centr} numerically on a grid of $120
\times 120$ points (for which convergence is reached) for a system of
size $L=1\mu$m, $\sigma=2$meV and $\ell_c=166$\AA.
From the evaluated eigenvalues $\varepsilon_\alpha$ and eigenstates
$\Phi_{\alpha,\vect{p}}$, we can derive the excitonic density of
states $\text{DoS} (\varepsilon)$, the coupling strength
$g_{\alpha,\vect{p}}$ and its squared average (Fig.~\ref{fig:gepsi}):
\begin{equation}
  g^2(\varepsilon,|\vect{p}|) = \Frac{1}{\text{DoS}
  (\varepsilon)}\langle \sum_{\alpha} |g_{\alpha,\vect{p}}|^2
  \delta(\varepsilon - \varepsilon_\alpha)\rangle \; ,
\label{eq:avgep}
\end{equation}
where $\langle \dots \rangle$ is the average over different disorder
realisations.  This quantity is related to the excitonic optical
density by $D(\varepsilon) = \text{DoS} (\varepsilon)
g^2(\varepsilon,0)$.

We now consider the following Hamiltonian describing excitons with
random energies $\varepsilon_{\alpha}$ dipole coupled via $g_{\alpha ,
\vect{p}}$ to the cavity field $\psi_{\vect{p}}$:
\begin{multline}
  \hat{H} = \sum_{\alpha} \Frac{\varepsilon_{\alpha}}{2}
  \left(b_{\alpha}^\dag b_{\alpha} + a_{\alpha} a_{\alpha}^\dag\right)
  + \sum_{\vect{p}} \omega_{\vect{p}} \psi_{\vect{p}}^\dag
  \psi_{\vect{p}}
\\ 
  + \Frac{1}{\sqrt{L^2}} \sum_{\alpha} \sum_{\vect{p}} \left(g_{\alpha
  , \vect{p}} \psi_{\vect{p}} b_{\alpha}^\dag a_{\alpha} +
  \text{h.c.}\right) \; .
\label{eq:hamil}
\end{multline}
Interactions are approximated by exclusion, so each exciton level
$\varepsilon_{\alpha}$ is modelled by an electron-hole pair $a$ and
$b$, with the total occupation restricted to one,
i.e. $b_{\alpha}^\dag b_{\alpha} + a_{\alpha}^\dag a_{\alpha} =
1$. For the thermodynamical properties of this model (e.g., the
critical temperature), this is a good assumption at low enough
densities, where only the strongly localised (Lifshitz) states in the
tail are populated. The dispersion for photons in a microcavity of
width $L_w$ is approximated as parabolic, $\omega_{\vect{p}} \simeq
\omega_0 + \vect{p^2}/2 m_{\text{ph}}$, where $\omega_0 = 2 \pi c /L_w
\sqrt{\epsilon}$ and $m_{\text{ph}} = 2\pi \sqrt{\epsilon}/c L_w$.  It
is convenient to rescale the exciton-light coupling $g_{\alpha ,
\vect{p}}$ according to $g_{\alpha , \vect{p}} \mapsto g_{\alpha ,
\vect{p}} \sqrt{\Ryx m_{\ex}/2\pi}$, where $N = \Ryx L^2 m_{\ex}/2\pi$
is the inverse level spacing measured in units of the excitonic
Rydberg energy. This corresponds to measuring the density of particles
in units of the Bohr radius squared. In thermal equilibrium, the total
number of excitations, $\hat{N} = \sum_{\alpha} (b_{\alpha}^\dag
b_{\alpha} + a_{\alpha} a_{\alpha}^\dag)/2 + \sum_{\vect{p}}
\psi_{\vect{p}}^\dag \psi_{\vect{p}}$, can be fixed by introducing a
chemical potential, $\mu$.  The dimensionless density $\rho \equiv
\langle \hat{N} \rangle /N$ corresponds to $(\langle
\hat{N}\rangle/L^2) a_{\ex}^2\, 4\pi \mu_r/m_{\ex}$, where $\langle
\hat{N}\rangle/L^2$ is the density of particles per area.

\begin{figure}
\begin{center}
\includegraphics[width=1\linewidth,angle=0]{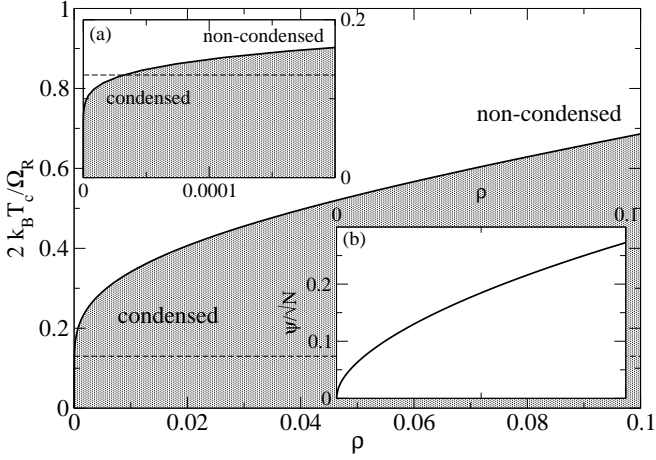}
\end{center}
\caption{\small Mean field phase diagram for the dimensionless
         critical temperature $2 T_c k_B/\Omega_{\mathrm{R}}$ versus
         the dimensionless density $\rho$ for effective zero detuning
         $\omega_0 - E_{\ex} = -0.94$meV and
         $\Omega_{\mathrm{R}}=26$meV.  A detail of the low density
         region is shown in the inset (a), while the plot of the
         mean-field order parameter $\psi/\sqrt{N}$ versus the density
         for $T k_{B} = 20$K ($2 T k_{B}/\Omega_{\mathrm{R}}=0.13$) is
         shown in the inset (b).}
\label{fig:criti}
\end{figure}
Making use of standard path integral techniques,
%
%
the mean-field equations for the static and uniform photon field
$\psi$ and for the total number of excitations can be simultaneously
solved in a similar way as in Refs.~\cite{paul,jonathan}. However, in
contrast to previous work, we introduce here a realistic description
for the excitons in the disordered quantum wells and make use of the
energies and coupling strengths evaluated numerically. In addition,
the average over different disorder realisations outside the energy
interval evaluated numerically (Fig.~\ref{fig:gepsi}(a)) is taken by
extrapolating the numerics: in the low energy Lifshitz tail, we
approximate the distribution of $|g_{\alpha,0}|^2$ with a delta
function at its extrapolated mean value $g^2 (\varepsilon,0)$, while,
in the high energy region, we use the Porter-Thomas distribution
$\mathcal{P} (x=|g_{\alpha,0}|^2) = \exp[-x/(2 \bar{x})]/\sqrt{2 \pi x
\bar{x}}$, where $\bar{x}=g^2 (\varepsilon,0)$. The scale of
$|g_{\alpha,0}|^2$ is set so as to fix the integrated optical density
$\int d\varepsilon D (\varepsilon)=\Omega_{\mathrm{R}}^2/4$, where
$\Omega_{\mathrm{R}}$ is the measured Rabi splitting.
In the resulting mean-field phase diagram (Fig.~\ref{fig:criti}),
increasing the density of particles at a given temperature, the system
goes under a phase transition from a non-condensed phase ($\psi=0$) to
a phase where the polaritons condense in the lowest momentum state
($\psi \ne 0$).

\begin{figure}
\begin{center}
\includegraphics[width=1\linewidth,angle=0]{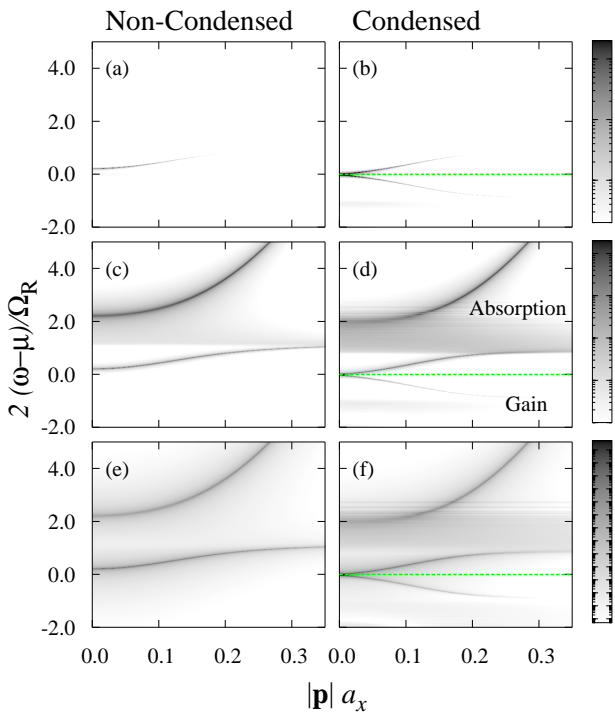}
\end{center}
\caption{\small Contourplot of the incoherent PL $P (\omega ,
	\vect{p})$ (a,b), the spectral weight $W(\omega , \vect{p})$
	(c,d), and the disorder averaged RRS intensity $\langle
	I_{\vect{p} \vect{q}} (\omega)\rangle$ for $|\vect{p}| =
	|\vect{q}|$ (e,f) versus the rescaled energy $2(\omega -
	\mu)/\Omega_{\mathrm{R}}$ and the dimensionless momentum
	$|\vect{p}| a_{\ex}$, for $\omega_0 - E_{\ex} = -0.94$meV,
	$\Omega_{\mathrm{R}}=26$meV, $k_{B} T = 20$K and
	$m_{\text{ph}} \Omega_{\mathrm{R}} a_{\ex}^2/2 = 0.01$:
	non-condensed (left column) ($\rho\simeq 0$, $\mu = -16.5$meV)
	and condensed (right column) ($\rho=3.6 \times 10^{-3}$, $\mu
	= -11.3$meV).}
\label{fig:rayle}
\end{figure}

The incoherent PL spectrum and the RRS response are evaluated by
considering fluctuations of the photonic field above the mean-field
solution:
%
\begin{gather}
\nonumber \delta \mathcal{S} \simeq \Frac{1}{2 k_B T} \sum_{\omega_h ,
  \vect{p}, \vect{q}}
  \begin{pmatrix} \delta \psi_{\omega_h , \vect{p}}^*\\ \delta
  \psi_{-\omega_h , -\vect{p}} \end{pmatrix}^{T}
  \mathcal{G}^{-1}_{\vect{p}\vect{q}} (\omega_h) \begin{pmatrix}
  \delta \psi_{\omega_h , \vect{q}} \\ \delta \psi_{-\omega_h ,
  -\vect{q}}^* \end{pmatrix}
\\ 
  \mathcal{G}^{-1}_{\vect{p} \vect{q}} (\omega_h) = \begin{pmatrix}
  K^{(1)}_{\vect{p} \vect{q}} (\omega_h) & K^{(2)}_{\vect{p} \vect{q}}
  (\omega_h)\\ K^{(2)}_{\vect{p} \vect{q}} (\omega_h) & {K^{(1)
  *}_{\vect{q} \vect{p}}} (\omega_h)
  \end{pmatrix} \; .
\end{gather}
Here, when $\omega_h = 2 \pi k_B T h \ne 0$, the matrix elements of
the inverse quasi-particle Green's function are given by
\begin{align}
\label{eq:kern1}
  K^{(1)}_{\vect{p} \vect{q}} (\omega_h)&= \delta_{\vect{p},\vect{q}}
  \left(i \omega_h + \tilde{\omega}_{\vect{p}}\right) + \Frac{1}{N}
  \sum_{\alpha} g_{\alpha, \vect{p}}^* g_{\alpha, \vect{q}} \\
  \nonumber &\times \left[i \omega_h \tilde{\varepsilon}_{\alpha}/2 -
  E_{\alpha}^2 -(\tilde{\varepsilon}_{\alpha}/2)^2\right]
  h(E_{\alpha})
\\
  K^{(2)}_{\vect{p} \vect{q}} (\omega_h) &= \Frac{\psi^2}{N}
  \Frac{1}{N} \sum_{\alpha} |g_{\alpha,0}|^2 g_{\alpha, \vect{p}}^*
  g_{\alpha, \vect{q}} h (E_{\alpha}) \; ,
\label{eq:kern2}
\end{align}
where $h(x)=\tanh (x/k_B T)/[x (\omega_h^2 + 4 x^2)]$, $E_{\alpha} =
\sqrt{(\tilde{\varepsilon}_{\alpha}/2)^2 + |g_{\alpha,0}|^2 \psi^2/N}$
is the energy of an exciton in a coherent field,
$\tilde{\varepsilon}_{\alpha} = \varepsilon_{\alpha} - \mu$ and
$\tilde{\omega}_{\vect{p}} = \omega_{\vect{p}} - \mu$. The
quasi-particle Green's function can be decomposed into (momentum)
diagonal and off-diagonal contributions, $K_{\vect{p}
\vect{q}}^{(1,2)} = K_{\vect{p} \vect{q}}^{(1,2)d}\delta_{\vect{p}
\vect{q}} + K_{\vect{p} \vect{q}}^{(1,2)o}$. The off-diagonal terms,
as they break translational invariance and therefore have a zero
average over different disorder realisations, can be neglected when
evaluating the incoherent PL intensity~\cite{whittaker}, $P (\omega ,
\vect{p}) = n_{B}(\omega) W(\omega,\vect{p})$, where $n_B(\omega)$ is
the Bose occupation factor and

\begin{equation}
   W(\omega,\vect{p})= \left. 2 \Im \mathcal{G}^{11}_{\vect{p}
  \vect{p}} (\omega_h) \right|_{i\omega_h = -\omega -i\eta} \; ,
\label{eq:specw}
\end{equation}
is the spectral weight. However, allowing the normal-modes supported
by the cavity to scatter via their excitonic component in directions
different from the incoming one, these terms are essential in finding
the RRS intensity:
\begin{equation}
  I_{\vect{p} \vect{q}} (\omega) = \left.  |\mathcal{G}^{11}_{\vect{p}
  \vect{q}} (\omega_h)|^2 \right|_{i\omega_h = -\omega -i\eta} \simeq
  F_{\vect{p}} S_{\vect{p} \vect{q}} F_{\vect{q}} \; .
\label{eq:RRSex}
\end{equation}
Here, the filter function $F_{\vect{p}}= ||K^{(1)}_{\vect{p}
\vect{p}}|^2 - [K^{(2)}_{\vect{p} \vect{p}}]^2|^{-2}$ describes the
propagation of the injected and detected photons via the normal modes
supported by the cavity, while the scattering function $S_{\vect{p}
\vect{q}}$
\begin{multline}
  S_{\vect{p} \vect{q}} = \left|K^{(1)o}_{\vect{p} \vect{q}} K^{(1)
  *}_{\vect{p} \vect{p}} K^{(1) *}_{\vect{q} \vect{q}} + K^{(1)o
  *}_{\vect{q} \vect{p}} K^{(2)}_{\vect{p} \vect{p}} K^{(2)}_{\vect{q}
  \vect{q}} \right.
\\
  \left. - K^{(2)o}_{\vect{p} \vect{q}} \left[K^{(2)}_{\vect{p}
  \vect{p}} K^{(1) *}_{\vect{q} \vect{q}} + K^{(1) *}_{\vect{p}
  \vect{p}} K^{(2)}_{\vect{q} \vect{q}}\right]\right|^2
\end{multline}
describes the probability to scatter, via the excitonic component,
from the injected momentum $\vect{p}$ to the detected one
$\vect{q}$. The incoherent PL $P (\omega , \vect{p})$ and spectral
weight $W(\omega,\vect{p})$, as also the disorder averaged RRS
response $\langle I_{\vect{p} \vect{q}} (\omega)\rangle$ are shown in
Fig.~\ref{fig:rayle} for two values of the density at a fixed
temperature, showing both the non-condensed
(Figs.~\ref{fig:rayle}(a,c,e)) and condensed case
(Figs.~\ref{fig:rayle}(b,d,f)).

The RRS signal reflects many of the features already present in the
spectral weight. The similarity is due to the filter functions, which
limit the RRS response to the normal modes supported by the cavity and
are responsible for the ring-shaped emission observed in
experiments~\cite{freixanet,langbein}. When uncondensed ($\psi \to
0$), $K^{(2)}=0$, the filter terms $|K^{(1)d}|^{-2}$ coincide with the
squared polariton Green's function and the scattering term
$|K^{(1)o}|^2$ gives the probability for an exciton to scatter from
$\vect{p}$ to $\vect{q}$. At ultralow densities, the model used here
is equivalent to that used in Ref.~\cite{whittaker} and the RRS
response is in agreement with that of
Refs.~\cite{whittakerRRS,shchegrov}. However, when condensed, the
polariton modes are replaced by new collective modes, the lower
polariton becomes a linear Goldstone mode, and two new branches appear
below the chemical potential, which are seen as gain in the spectral
weight. These changes are also seen in the RRS response, including RRS
response at energies below the chemical potential
(Figs.~\ref{fig:rayle}(f)).

These noticeable changes could be observed in both PL emission, the
product of the spectral weight times the Bose occupation factor
$n_B(\omega)$, and the RRS response. However, because of the
occupation factor, PL emission from above the chemical potential is
exponentially suppressed. In addition, the PL as it is plotted in
Fig.~\ref{fig:rayle} excludes the emission from the condensate, which
in experiments might obscure these features. In contrast, RRS response
is not weighted by occupation, but instead by the scattering function,
$S_{\vect{p} \vect{q}}$, which depends on the modulus squared of the
Green's function for an exciton in a coherent field. As a consequence,
RRS represents a unique probe for observing condensation in polariton
microcavities.

Using the full distribution of oscillator strengths has observable
effects in the condensed state and is vital in our treatment. When
uncondensed, while the spectral weight depends only on the excitonic
optical density (and thus on the average of $|g_{\alpha,0}|^2$), the
RRS depends also on $|g_{\alpha,0}|^4$. This determines a sharper
energy dependence in RRS than in optical density. However, when
condensed, both optical responses are determined by the the full
distribution of oscillator strengths, rather than only the mean
squared oscillator strength and its mean fourth power. This is because
the energy $E_{\alpha}$ of an exciton in the presence of a coherent
field leads to a dependence on the full distribution of
$|g_{\alpha,0}|$. The density of states of these excitonic
quasiparticles can be directly seen in the spectral weight between the
upper and lower polariton modes. With a constant $g$, the minimum of
$E_{\alpha}$ is sharp, occurring for excitons where
$\tilde{\varepsilon}_{\alpha}=0$. In contrast, with the distribution
of $|g_{\alpha,0}|$ used here, the minimum of $E_{\alpha}$ occurs for
states where there is first a finite probability of having negligible
coupling to light. Below this energy, other than the linear mode,
there is a suppression of the spectral weight. The energy dependence
of the distribution of coupling strengths is also important, and is
responsible for the decrease of polariton splitting at higher
densities. As the chemical potential increases, the excitons with
energies close to the chemical potential couple less strongly to
light, and the effective splitting decreases.

The treatment in this Letter considers only disorder acting on the
excitons; it has been suggested~\cite{langbein,gurioli} that disorder
of the mirrors, acting on the photons might also be relevant. The
inclusion of such disorder can be expected to modify the scattering
term, but the filter functions are expected to remain unaffected.

To conclude, we have described how polariton condensation in
semiconductor microcavities can be investigated by RRS. By making use
of a realistic model of disorder in quantum wells, we identified the
changes which can be observed in the RRS spectrum when, with
increasing density, the system, still in the strong coupling regime,
crosses the the phase boundary to a condensed state. Because RRS, in
contrast to PL, is not weighted by the Bose factor, and because RRS
does not contain a strong signal at the condensate frequency, which
could obscure the subtle features in PL, RRS provides a promising
probe of polariton condensation in semiconductor microcavities.

We are grateful to B. D. Simons, Le Si Dang, J. Kasprzak and W.
Langbein for suggestions and useful discussions. FMM and MHS would
like to acknowledge the financial support of EPSRC.
This work is supported by the EU Network ``Photon mediated phenomena
in semiconductor nanostructures'' HPRN-CT-2002-00298.

%


\begin{thebibliography}{99}
%

\bibitem{lesidang}
%
Le Si Dang \emph{et al.}, Phys. Rev. Lett. \textbf{81}, 3920 (1998).


\bibitem{deng} 
%
H. Deng \emph{et al.}, Science \textbf{298}, 199 (2002); H. Deng
\emph{et al.}, PNAS \textbf{100}, 15318 (2003).

\bibitem{richard}
%
M. Richard \emph{et al.}, Phys. Rev. Lett. \textbf{94}, 187401 (2005).

\bibitem{private}
%
Le Si Dang, J. Kasprzak, H. Deng, private commun.


\bibitem{jonathan} 
%
J. Keeling \emph{et al.}, Phys. Rev. Lett. \textbf{93}, 226403 (2004);
Phys. Rev. B \textbf{72}, 115320 (2005).  

\bibitem{bosonic}
%
V. Savona \emph{et al.}, Phase Transitions \textbf{68}, 169 (1999);
F. P. Laussy \emph{et al.}, Phys. Rev. Lett. \textbf{93}, 016402
(2004); A. Kavokin and G. Malpuech, \emph{Cavity polaritons}, in Thin
Films and Nanostructures vol. 32 (Elsevier, NY, 2003).

%
\bibitem{freixanet}
%
T. Freixanet \emph{et al.}, Phys. Rev. B \textbf{60}, R8509 (1999).

\bibitem{langbein}
%
W. Langbein and J. M. Hvam, Phys. Rev. Lett. \textbf{88}, 047401
(2002).

\bibitem{runge_review}
%
E. Runge, ``Excitons in Semiconductors Nanostructures'', in
\emph{Solid State Physics} vol. \textbf{57}, Academic Press, Amsterdam
(2002).

\bibitem{zimmermann}
%
R. Zimmermann, F. Grosse, and E. Runge, Pure and Appl. Chem.
\textbf{69}, 1179 (1997); E. Runge and R. Zimmermann,
Phys. Stat. Sol. (b) \textbf{221}, 269 (2000).

\bibitem{carusotto}
%
I. Carusotto and C. Ciuti, Phys. Rev. Lett. \textbf{93}, 166401
(2004); preprint cond-mat/0502585.


%

\bibitem{paul}
%
P. R. Eastham and P. B. Littlewood, Solid Sate Commun. \textbf{116},
357 (2000); Phys. Rev. B \textbf{64}, 235101 (2001).

\bibitem{whittaker}
%
D. M. Whittaker, Phys. Rev. Lett. \textbf{80}, 4791 (1998).

\bibitem{whittakerRRS}
%
D. M. Whittaker, Phys. Rev. B \textbf{61}, R2433 (2000).

\bibitem{shchegrov}
%
A. V. Shchegrov \emph{et al.}, Phys. Rev. Lett. \textbf{84}, 3478
(2000).

\bibitem{gurioli} 
%
M. Gurioli \emph{et al.}, Phys. Rev. Lett. \textbf{94}, 183901 (2005).


\end{thebibliography}
\end{document}